\documentclass[aps,prd,onecolumn,groupedaddress,nofootinbib,amssymb]{revtex4}

\usepackage{graphicx,bm,color}
\usepackage{amsmath}
\usepackage{amssymb}
\usepackage{amsfonts}
\usepackage{wrapfig}
\usepackage{cancel}
\usepackage{url}

\newcommand{\be}{\begin{equation}}
\newcommand{\ee}{\end{equation}}
\newcommand{\bea}{\begin{eqnarray}}
\newcommand{\eea}{\end{eqnarray}}
\newcommand{\beaa}{\begin{eqnarray*}}
\newcommand{\eeaa}{\end{eqnarray*}}

\newcommand{\nn}{\nonumber \\}

\allowdisplaybreaks[4]

\begin{document}

\def\theequation{\arabic{section}.\arabic{equation}}

\title{Effects of modified gravity on the turnaround radius in cosmology}

\author{Shin'ichi~Nojiri$^{1,2}$, 
Sergei~D.~Odintsov$^{3,4,5}$ and Valerio~Faraoni$^{6}$,
}
\affiliation{
$^1$Department of Physics, Nagoya University, Nagoya 464-8602, Japan\\
$^2$Kobayashi-Maskawa Institute for the Origin of Particles and the
Universe, Nagoya University, Nagoya 464-8602, Japan\\
$^3$Instituci\'{o} Catalana de Recerca i Estudis Avan\c{c}ats
(ICREA), Passeig Llu\'{i}s Companys, 23, 08010 Barcelona, Spain\\
$^4$ Institute of Space Sciences (ICE, CSIC) C. Can Magrans s/n, 08193
Barcelona, Spain \\
$^5$ Institute of Space Sciences of Catalonia (IEEC), Barcelona, Spain\\
$^6$Department of Physics and Astronomy and STAR Research Cluster, Bishop's 
University, 2600 College Street, Sherbrooke, Qu\'ebec, Canada J1M 1Z7
}

\begin{abstract}

We revisit the concept of turnaround radius in cosmology, in the
context of modified gravity. While preliminary analyses were limited to
scalar-tensor/$F(R)$ gravity, we extend the definition and the study of
this quantity to a much broader class of theories including also quantum $R^2$ 
gravity. The turnaround radius
is computed in terms of the parameters of the theory and it is shown that
a deviation not larger than 10\% of this quantity from its value in
Einstein's theory could constrain the model parameters and even rule out
some current theories.  

\end{abstract}
\maketitle

\section{Introduction}
\label{sec:1}
\setcounter{equation}{0}

Modifying general relativity (GR) is a necessity from the
theoretical physics point of view. In fact,
virtually all attempts to quantize GR
modify the Einstein-Hilbert action by adding extra
dynamical fields or non-local terms, or by introducing higher order
derivatives in the field equations (see \cite{Buchbinder:1992rb} for review). These 
corrections
are not necessarily Planck-scale suppressed. For example, the simplest
string theory, the bosonic string theory, reduces
to an $\omega=-1$ Brans-Dicke gravity in its  low-energy limit
\cite{Callan:1985ia,Fradkin:1985ys} and the $F(R)$ theories of gravity which are
nowadays popular to
explain away dark energy are nothing but scalar-tensor theories in
disguise \cite{Capozziello:2011et,Cai:2015emx,Nojiri:2006ri,Sotiriou:2008rp,
Nojiri:2017ncd,Nojiri:2010wj,Clifton:2011jh}. 

A significant body of experimental efforts aiming to test
gravity at all
possible astrophysical and cosmological scales has emerged in the last
decade.  There is little doubt, however,  that the main motivation to
question GR comes from cosmology. The present acceleration of the
universe discovered in 1998 with
type Ia supernovae requires an explanation. While a cosmological
constant $\Lambda$ offers a possible explanation in principle, it is
peppered with enormous fine-tuning problems, which has led to the
introduction of the
completely {\em ad  hoc} concept of dark energy  (see
\cite{Bamba:2012cp} for a review). Many authors, dissatisfied
with these approaches, have turned to the
possibility of modifying gravity at large scales (\cite{Capozziello:2003tk,Nojiri:2003ft}, see
also \cite{Capozziello:2011et,Cai:2015emx,Nojiri:2006ri,Sotiriou:2008rp, Nojiri:2017ncd, Nojiri:2010wj,
Clifton:2011jh,Baker:2014zba} for reviews). Modulo some
fine-tuning, the
idea works in principle, but many modified gravity
models (and many dark energy models as well) fit the observational data.
Therefore, one would like to avail oneself  of all the tests of gravity
which become available, at all scales and in all regimes, to obtain the
correct scenario. In this context, the turnaround radius may be useful.

{The concept of turnaround radius has been around for many 
years under various names (see, {\em e.g.}, 
\cite{Stuchlik1, Stuchlik:1999qk, Stuchlik3, 
Stuchlik:2008dv,Mizony:2004sh,Stuchlik:2011zz,Roupas:2013xga,Nolan:2014maa}): 
radius of the ``zero velocity surface'' or of the ``effective 
sphere of influence of a cosmic structure'', ``zero gravity radius'', 
``critical radius'', ``maximum size of large scale structures'', 
``maximum size of bound cosmic structures'', and 
``maximum turnaround radius''. The literature seems to have settled on the term 
``turnaround radius'', which we adopt. Its study has emerged 
only recently as a possible way to test dark 
energy in GR by comparing theoretical predictions with astronomical 
observations \cite{Busha:2003sz,Pavlidou:2013zha,Pavlidou:2014aia}.

In an
accelerating Friedmann-Lem\^aitre-Robertson-Walker (FLRW) universe, there
is a maximum physical (areal) radius, called {\em turnaround radius}
$r_\mathrm{TA}$ such
that any spherical shell of dust (composed of test particles following
radial time-like geodesics) located outside $r_\mathrm{TA}$ and given zero 
radial
velocity initially, cannot collapse but is forced to expand forever by the
cosmic acceleration. A similar dust shell located inside the
turnaround radius, instead, will collapse. The turnaround radius
constitutes the maximum possible radius of a bound structure in an
accelerating FLRW universe. Early comparisons of the theoretical
turnaround radius in the GR-based $\Lambda$CDM model with celestial
objects have been carried out \cite{Pavlidou:2013zha,Pavlidou:2014aia,Busha:2003sz} but the
astronomical error is quite large. Nevertheless, the method is quite
promising in principle. Within GR, the concept of turnaround radius has
been made precise,
more rigorous, and gauge-invariant (to first order in the perturbations of
an exact FLRW cosmos) by using an approach \cite{Faraoni:2015saa, Venice2017} 
based on the Hawking-Hayward
quasi-local energy \cite{Hawking:1968qt,Hayward:1993ph,Hayward:1994bu}. 
The numerical
value of the turnaround radius estimated in this way, however, turns out
to be quite close to that estimated with the previous method
\cite{Faraoni:2015saa, Venice2017}. The definition of quasilocal
energy in GR is not unique (see \cite{Szabados:2009eka} for a review), but it is
reeassuring that the Hawking-Hayward construct and the Brown-York
quasilocal mass (in an appropriate gauge) provide the same answer to first
order in the cosmological perturbations \cite{Lapierre-Leonard:2017pvj} 
(higher order calculations are futile in view of the large observational 
errors in the determination of the turnaround radius).

Even  more interesting is the fact that the turnaround radius can, in
principle, be used to discriminate between GR and alternative theories of
gravity. Preliminary analyses of the turnaround radius in
scalar-tensor and $F(R)$ gravity and others were performed in
Refs.~\cite{Faraoni:2015zqa, Capozziello:2018oiw,Lopes:2018uhq,Bhattacharya:2016vur}.
Unfortunately, the status of quasilocal energy (which is already non-unique in GR 
\cite{Szabados:2009eka}), is not clear in modified gravity, in spite of some attempts to 
generalize this definition within the restricted context of scalar-tensor theories 
\cite{Cai:2009qf,Cai:2008mh,Wu:2007se,Cognola:2011nj,Faraoni:2015sja,Hammad:2016yjq}. Therefore, 
in modified gravity one is forced, at least for the moment, to
give up the quasilocal energy approach and to pursue other approaches.
Recently, it was reported that the upper
bound set by GR on  the turnaround radius is significantly exceeded  in the
galaxy group NGC 5353/4 \cite{Lee:2015upn,Lee:2016qpt}. The need to take into account 
 the error introduced by the non-sphericity of the system has been emphasized \cite{Lee:2016oyu}, 
together with the fact that one should expect a distribution of the value of the turnaround radius 
among different astronomical systems and, therefore, an excess in this quantity would be significant 
from the statistical point of view rather than for individual systems \cite{Lee:2016bec, Lee:2017ejv}. Currently, the 
observational search is focussing on galaxy groups with web-like structures in their neighbouring zones, 
and six more groups exceeding the general-relativistic prediction for the turnaround radius have 
been reported \cite{Lee:2017ejv}. In view of these very promising observational developments and of its 
potential consequences as a probe of the correct theory of gravity it is worth studying this subject 
more in depth.

In GR and in an asymptotically de Sitter spacetime the turnaround radius depends on the 
cosmological constant, the gravitational coupling, and the mass contained inside this radius. It 
is important to realize that, in modified gravity, both the effective cosmological constant and 
the gravitational coupling are changed from their GR values 
\cite{Faraoni:2015zqa,Faraoni:2015saa}.  Therefore, by comparing the turnaround radius in 
modified gravity theories with the size of large scale structures, we constrain the modified 
gravity theory.

Lacking a clear concept of quasilocal energy when we leave the GR context, as already noted, we 
resort to a different definition of turnaround radius than the one of \cite{Faraoni:2015saa}. 
The idea is that, at the turning radius, the gravitational force balances the inertial force 
generated by the accelerating expansion of the universe. Here, we consider the generalization of 
the turnaround radius for $F(R)$ gravity and $R+R^2+R_{\mu\nu}R^{\mu\nu}$ gravity and its 
generalizations (like one-loop corrected quantum $R^2$ gravity). In 
Ref.~\cite{Capozziello:2018oiw}, only the asymptotically de Sitter spacetime background was 
considered, where the effective cosmological constant and the effective gravitational coupling 
are constant. In this paper instead, we consider power-law expansion in $F(R)$ gravity, where 
the effective gravitational ``constant'' is time-dependent and therefore the expression of the 
turnaround radius changes from that in GR coupled with a cosmological fluid. Even in the case of 
$R+R^2+R_{\mu\nu} R^{\mu\nu}$ gravity, which includes the square of the Ricci tensor introducing 
new degrees of freedom, the Schwarzschild-de Sitter spacetime is an exact solution. We 
investigate the possible observational constraints on the parameters of the models coming from 
the turnaround radius. A problem arising in these models is that we observe the effective 
coupling constant as defined by Newton's law and, therefore, it is difficult to distinguish the 
modified gravity theory from Einstein gravity in the de Sitter spacetime background using the 
turnaround radius.  This fact tells us that we need to find the coupling constant in the 
Einstein-Hilbert term with independent methods.

The plan of this paper is as follows. In the next section, we review the turnaround radius and 
discuss the case of $F(R)$ gravity, especially in a power-law expanding universe. 
Sec.~\ref{sec:3} derives general formulas for a broad class of gravitational theories, while 
Sec.~\ref{sec:4} focuses on a particular model of $R^2$ gravity including Ricci-squared term.  
Sec.~\ref{sec:5} constrains a yet more general class of models which represent one-loop 
corrected $R^2$ gravity and Sec.~\ref{sec:6} contains a discussion and the conclusions. We use 
the metric signature $-+++$ and units in which the speed of light $c$ assumes the value unity.  
$G_N$ is Newton's constant and otherwise we follow the notation of Ref.~\cite{Wald:1984rg}.

\section{Generalization of turnaround radius}
\label{sec:2}
\setcounter{equation}{0}

For the spherical and (locally) static Schwarzschild-like metric 
written in curvature coordinates 
\begin{equation}
\label{TuA01}
ds^2 = - A(r) dt^2 + \frac{1}{A(r)} dr^2 + r^2 d\Omega_{(2)}^2 \, ,
\end{equation}
where $d\Omega_{(2)}^2=d\theta^2 +\sin^2 \theta \, d\varphi^2$ is
the line element on the unit 2-sphere, the turnaround radius
$r_\mathrm{TA}$ (an areal radius) is defined by $r=r_\mathrm{TA}$
which satisfies the condition \cite{Capozziello:2018oiw}
\begin{equation}
\label{TuA02}
0 = A' \left( r_\mathrm{TA} \right) \, .
\end{equation}
This is because $A(r)$ is related with the effective 
gravitational potential $\phi$ by 
\begin{equation}
\label{TuA03}
A(r) = 1 + 2 \phi\, .
\end{equation}
In particular,  in the case of the Schwarzschild-de Sitter 
spacetime, 
\begin{equation}
\label{TuA04}
A(r) = 1 - \frac{2G_\mathrm{N}M}{r} - \frac{r^2}{l^2} \, ,
\end{equation}
with Newton's gravitational constant $G_\mathrm{N}$,
the de Sitter length parameter $l$, and the mass $M$ of the 
gravitational source,  we find 
\cite{Pavlidou:2013zha, Pavlidou:2014aia} 
\begin{equation}
\label{TuA05}
r_\mathrm{TA}^3 = G_\mathrm{N}M l^2 \, .
\end{equation}
Following previous literature, we discuss a spherical
inhomogeneity embedded in a spatially flat FLRW
background universe with line element
\begin{equation}
ds^2 =-dt^2 +a^2(t) \left( dr^2 +r^2 d\Omega_{(2)}^2 \right) \, ,
\end{equation}
and scale factor $a(t)$. 
The turnaround radius can be regarded as 
the radius where the gravitational force
\begin{equation}
\label{TuA06}
F_\mathrm{g} = \frac{G_\mathrm{N}mM}{r^2}
\end{equation}
acting on a test mass $m$ or an observer, balances the inertial
force generated by the expansion of the universe and discussed for
the Big Rip \cite{Caldwell:2003vq}
and the Little Rip \cite{Frampton:2011sp,Frampton:2011rh}
\begin{equation}
\label{TuA07}
F_\mathrm{in} = r m \frac{\ddot a}{a} = r m \left( \dot H+ H^2 \right) \, .
\end{equation}
Here $H$ is the  Hubble rate, $H\equiv \dot a/a$ 
where an overdot denotes differentiation
with respect to the comoving time $t$ of the FLRW background. 
In the case of the de Sitter universe, where $H=1/l$, the
equation expressing 
the balance $F_\mathrm{g} = F_\mathrm{in}$ reproduces the 
well known result of Eq.~(\ref{TuA05}). 
In a more general expanding universe, we obtain the 
following expression of $r_\mathrm{TA}$
\begin{equation}
\label{TuA08}
r_\mathrm{TA}^3 = \frac{G_\mathrm{N}M}{\dot H + H^2} \, .
\end{equation}
Equation~(\ref{TuA08}) can be used, for example, 
in the universe with the power law expansion. 
This criterion for the turnaround radius is conceptually different from
previous definitions given in the literature
\cite{Pavlidou:2013zha, Pavlidou:2014aia, Faraoni:2015saa} in the context
of GR, although the numerical value of this quantity
can be numerically close to that computed with other definitions in some
physically interesting situations. 

As is clear from the expression of the inertial force~(\ref{TuA07}),
$F_\mathrm{in}$ is repulsive in an accelerating 
expanding universe, $\ddot a>0$, but in a decelerating 
expanding universe, $\ddot a<0$, as in the
matter/radiation-dominated universe, the inertial force
$F_\mathrm{in}$ becomes attractive, 
the turnaround radius $r_\mathrm{TA}$ does not exist, and we do not
obtain any constraint.  Even in an accelerating universe, the
inertial force becomes smaller as time passes if the
effective Equation of State (EoS) parameter $w \equiv P/\rho$ 
(where $\rho$ and $P$ are the energy density and pressure of the
cosmic fluid, respectively) is larger than $-1$, 
{\em i.e.}, for $-1<w < - 1/3$ and the turnaround radius becomes larger.
An interesting point in $F(R)$ gravity is that the effective 
gravitational coupling $G_\mathrm{eff} \propto 1/F'(R)$ is
time-dependent. For example, if $F(R)$ behaves as $F(R)\sim R^\alpha$ with
a constant $\alpha$, $F'(R) \propto R^{\alpha -1} \propto t^{-2 \left(
\alpha - 1 \right)}$ because $R =6\left( \dot{H}+2H^2 \right) $
behaves as $R\propto t^{-2}$. Therefore, we find 
\begin{equation}
\label{TuA09}
r_\mathrm{TA}^3 \propto t^{ 2(\alpha -2)}\, .
\end{equation}
Then, if $\alpha>2$, the turnaround radius $r_\mathrm{TA}$ becomes larger
as time passes but becomes smaller if $\alpha<2$.
We should note that if $F(R)$  behaves as $F(R)\sim R^\alpha$, the scale
factor $a$ behaves as \cite{Nojiri:2003ft,Nojiri:2006ri}
\begin{equation}
\label{F1}
a (t) \propto t^{\frac{(\alpha - 1)(2 \alpha - 1)}{\alpha -2}}\, ,
\end{equation}
when we neglect the contribution from the matter, 
therefore $\alpha>2$ corresponds to a 
phantom universe and $1<\alpha<2$ to a quintessence one. 
A general $\alpha$ corresponds to Einstein gravity coupled
with a perfect fluid with the effective equation of state parameter, 
\begin{equation}
\label{FF1}
w_\mathrm{eff} = - \frac{ \left( 6\alpha^2 - 11 \alpha +7 \right)}{3
\left(\alpha - 1\right) \left( 2\alpha - 1 \right)} \, .
\end{equation}
In the case that matter with the EoS parameter $w$ couples
with $F(R)$ gravity, the effective EoS parameter is 
\begin{equation}
\label{FF2}
w_\mathrm{eff} = - 1 + \frac{1+w}{\alpha} \, .
\end{equation}
In both cases~(\ref{FF1}) or~(\ref{FF2}), if
$w_\mathrm{eff}< - 1/3$, the inertial force  (\ref{TuA07}) becomes
repulsive and there appears the turnaround radius~(\ref{TuA08}).
We should note that, even if the expansion of the universe is identical,
the behavior of the turnaround radius is different in 
$F(R)$ gravity and in Einstein gravity coupled with a perfect fluid.
When Einstein gravity couples with only one kind of perfect  fluid with a 
constant EoS  parameter $w \neq -1$, the Hubble rate $H$ always
behaves as $H\propto t^{-1}$.  Then, Eq.~(\ref{TuA08}) tells us that, in GR, 
\begin{equation}
\label{FF2B}
r_\mathrm{TA}^3 \propto t^2\, ,
\end{equation}
which is different from the expression~(\ref{TuA09}) of $F(R)$ gravity. 
If $\alpha>6$, the turnaround radius $r_\mathrm{TA}$ in 
$F(R)$ gravity is larger than the corresponding radius~(\ref{FF2B}) in 
Einstein gravity and, therefore, comparatively larger bound structures can 
form more easily in the universe. On the other hand, if $\alpha<6$, the
turnaround radius $r_\mathrm{TA}$ in $F(R)$ gravity
is smaller than the radius~(\ref{FF2B}) in Einstein gravity and the size of 
bound structures in the universe becomes smaller.

The concept of the inertial force~(\ref{TuA07}) generated by the
expansion of the universe has been introduced in the investigation of the
Little Rip~\cite{Frampton:2011sp,Frampton:2011rh}. In the Little Rip
scenario, the Hubble rate $H$ goes to infinity in the infinite future $t\to +\infty$, 
while $H$ diverges at a finite future $t\to t_s$ in the Big Rip scenario.
Anyway, as $H$ becomes larger and larger, which means $\dot H>0$,
the turnaround radius~(\ref{TuA08}) becomes smaller and smaller.
In Eq.~(\ref{TuA08}), we have considered the balance between 
the Newtonian force and the inertial force~(\ref{TuA07}).  If the radius becomes 
of the order of the human size, say, the electromagnetic force between 
molecules becomes stronger than the gravitational force, 
and then we need to consider the balance between the
electromagnetic force and the inertial force. If the size of
the turnaround radius becomes of the order of the nuclear size, 
we further need to consider the  balance between nuclear and inertial forces.

\section{A class of models}
\label{sec:3}
\setcounter{equation}{0}

We now consider the following class of  models of gravity,
\begin{equation}
\label{TuA21B}
S = \frac{1}{2\kappa^2} \int d^4 x \, \sqrt{ - g } \Big[ F(R) + G(R)
R_{\mu\nu} R^{\mu\nu} \Big] + S_\mathrm{matter} \,.
\end{equation}
By varying the action~(\ref{TuA21B}) with respect to the (inverse) metric 
$g^{\mu\nu}$, one obtains the fourth order field equations 
\begin{align}
\label{TuA22B}
0 = & \frac{1}{2} \, g_{\mu\nu} \left[ F(R) + G(R) R_{\rho\sigma}
R^{\rho\sigma}  \right] - 2 G(R) R_\mu^{\ \rho} R_{\nu\rho}
  - \left[ F'(R) + G'(R) R_{\rho\sigma} R^{\rho\sigma}  \right]
R_{\mu\nu} \nn
&\nn
& + \left( \nabla_\mu \nabla_\nu - g_{\mu\nu} \nabla^2 \right) \left[
F'(R) + G'(R) R_{\rho\sigma} R^{\rho\sigma}  \right] \nn
&\nn
& + \nabla_\mu \nabla^\rho \left[ G(R) R_{\rho\nu}\right) + \nabla_\nu
\nabla^\rho \left(G(R) R_{\rho\mu} \right] \nn
&\nn
& - \nabla^2 \left[ G(R) R_{\mu\nu} \right] - g_{\mu\nu}  \nabla^\rho
\nabla^\sigma \left[ G(R) R_{\rho\sigma} \right] + \kappa^2 T_{\mu\nu}\, .
\end{align}
Here we have used the following formulas,
\begin{align}
\label{curvature}
\delta R_{\mu\nu} =& \frac{1}{2}\left[\nabla^\rho\left(\nabla_\mu \delta g_{\nu\rho} 
+ \nabla_\nu \delta g_{\mu\rho}\right) - \nabla^2 \delta g_{\mu\nu} 
 - \nabla_\mu \nabla_\nu \left(g^{\rho\lambda}\delta g_{\rho\lambda}\right)\right] \, ,\\
& \ \nn
\label{curvature2}
\delta R =& -\delta g_{\mu\nu} R^{\mu\nu} + \nabla^\mu \nabla^\nu \delta g_{\mu\nu} 
 - \nabla^2 \left(g^{\mu\nu}\delta g_{\mu\nu}\right)\, . 
\end{align}

When $T_{\mu\nu}=0$, if we assume that the scalar curvature 
and the Ricci tensor are covariantly constant,
\begin{equation}
\label{TuA5B}
R = \frac{12}{l^2}\, , \quad \quad R_{\mu\nu} = \frac{3}{l^2} \,
g_{\mu\nu} \, ,
\end{equation}
we obtain the algebraic equation for $1/l^2$,
\begin{align}
\label{TuA22}
0 = \frac{1}{2} \, F(R_0) - \left[  F'(R_0) + \frac{36G'(R_0)}{l^4}
\right]   \frac{3}{l^2}\, , \quad\quad  R_0 \equiv \frac{12}{l^2} \, .
\end{align}
If a real positive solution $1/l^2$ exists, the de Sitter 
and the Schwarzschild-de Sitter spacetimes 
(\ref{TuA04}) are solutions of Eq.~(\ref{TuA22B}), with 
\begin{equation}
\label{TuA8}
A(r) = 1 - \frac{2G_\mathrm{eff}M}{r} - \frac{r^2}{l^2} \, ,
\end{equation}
except for the fact that Newton's gravitational constant $G_\mathrm{N}$
is now replaced by the effective one $G_\mathrm{eff}$.

In order to define the effective gravitational constant $G_\mathrm{eff}$,
we consider the perturbation of 
Eq.~(\ref{TuA22B}),
\begin{equation}
\label{TuA23}
g_{\mu\nu} \to g_{\mu\nu} + h_{\mu\nu} \, .
\end{equation}
Because
\begin{equation}
\label{TuA9bB}
\delta\Gamma^\kappa_{\mu\nu} =\frac{1}{2}g^{\kappa\lambda}\left(
\nabla_\mu h_{\nu\lambda} + \nabla_\nu h_{\mu\lambda}
  - \nabla_\lambda h_{\mu\nu} \right)\, ,
\end{equation}
we obtain
\begin{align}
\label{TuA10B}
\delta R =& - h_{\mu\nu} R^{\mu\nu} + \nabla^\mu \nabla^\nu h_{\mu\nu}
  - \nabla^2 \left(g^{\mu\nu} h_{\mu\nu}\right)\, , \nn
&\nn
\delta R_{\mu\nu} = &  \frac{1}{2}\left[\nabla^\rho\left(\nabla_\mu
h_{\nu\rho} + \nabla_\nu h_{\mu\rho}\right) - \nabla^2 h_{\mu\nu}
- \nabla_\mu \nabla_\nu \left(g^{\rho\lambda}
h_{\rho\lambda}\right)\right] \nn
&\nn
= & \frac{1}{2}\left[\nabla_\mu\nabla^\rho h_{\nu\rho}
+ \nabla_\nu \nabla^\rho h_{\mu\rho} - \nabla^2 h_{\mu\nu}
  - \nabla_\mu \nabla_\nu \left(g^{\rho\lambda} h_{\rho\lambda}\right)
  - 2R^{\lambda\ \rho}_{\ \nu\ \mu} h_{\lambda\rho}
+ R^\rho_{\ \mu} h_{\rho\nu} + R^\rho_{\ \mu} h_{\rho\nu} \right]\, .
\end{align}
Then we find
\begin{align}
\label{TuA24B}
0 = & \frac{1}{2} \, h_{\mu\nu} \left\{ F(R) + G(R) R_{\rho\sigma}
R^{\rho\sigma}  \right\}
+ \frac{1}{2} \, g_{\mu\nu} \left\{ F'(R) + G'(R) R_{\rho\sigma}
R^{\rho\sigma}  \right\}\left(
  - h_{\xi\eta} R^{\xi\eta} + \nabla^\xi \nabla^\eta h_{\xi\eta}
  - \nabla^2 \left(g^{\xi\eta} h_{\xi\eta}\right) \right) \nn
&\nn
& - 2 G'(R) R_\mu^{\ \rho} R_{\nu\rho} \left(
  - h_{\xi\eta} R^{\xi\eta} + \nabla^\xi \nabla^\eta h_{\xi\eta}
  - \nabla^2 \left(g^{\xi\eta} h_{\xi\eta}\right) \right) \nn
&\nn
&  - \frac{1}{2} \left\{ F'(R) + G'(R) R_{\rho\sigma} R^{\rho\sigma}
\right\}
\left\{\nabla^\rho\left(\nabla_\mu h_{\nu\rho} + \nabla_\nu
h_{\mu\rho}\right) - \nabla^2 h_{\mu\nu}  - \nabla_\mu \nabla_\nu
\left(g^{\rho\lambda} h_{\rho\lambda}\right)\right\} \nn
&\nn
&  - 2 G(R_0) R_\mu^{\ \eta} \left\{\nabla^\rho\left(\nabla_\eta h_{\nu\rho}
+ \nabla_\nu h_{\eta\rho}\right) - \nabla^2 h_{\eta\nu}   - \nabla_\eta
\nabla_\nu \left(g^{\rho\lambda} h_{\rho\lambda}\right)\right\} \nn
&\nn
&  - 2 G(R_0) R_\nu^{\ \eta} \left\{\nabla^\rho\left(\nabla_\eta h_{\mu\rho}
+ \nabla_\mu h_{\eta\rho}\right) - \nabla^2 h_{\eta\mu}
  - \nabla_\eta \nabla_\mu \left(g^{\rho\lambda}
h_{\rho\lambda}\right)\right\} \nn
&\nn
& - \left\{ F''(R) + G''(R) R_{\rho\sigma} R^{\rho\sigma}  \right\} R_{\mu\nu} 
\left(
- h_{\xi\eta} R^{\xi\eta} + \nabla^\xi \nabla^\eta h_{\xi\eta}
  - \nabla^2 \left(g^{\xi\eta} h_{\xi\eta}\right) \right) \nn
&\nn
& + \left( \nabla_\mu \nabla_\nu - g_{\mu\nu} \nabla^2 \right) \left\{ \left( 
F''(R) + G''(R) R_{\rho\sigma} R^{\rho\sigma}\right)\left(
  - h_{\xi\eta} R^{\xi\eta} + \nabla^\xi \nabla^\eta h_{\xi\eta}
  - \nabla^2 \left(g^{\xi\eta} h_{\xi\eta}\right) \right) \right\} \nn
&\nn
& + \nabla_\mu \nabla^\rho \left\{ G'(R) R_{\rho\nu} \left(
- h_{\xi\eta} R^{\xi\eta} + \nabla^\xi \nabla^\eta h_{\xi\eta}
  - \nabla^2 \left(g^{\xi\eta} h_{\xi\eta}\right) \right)
\right\} \nn
&\nn
& + \nabla_\nu \nabla^\rho \left\{G'(R) R_{\rho\mu} \left(
  - h_{\xi\eta} R^{\xi\eta} + \nabla^\xi \nabla^\eta h_{\xi\eta}
  - \nabla^2 \left(g^{\xi\eta} h_{\xi\eta}\right) \right)
\right\} \nn
&\nn
& - \nabla^2 \left\{ G'(R) R_{\mu\nu} \left(
- h_{\xi\eta} R^{\xi\eta} + \nabla^\xi \nabla^\eta h_{\xi\eta}
  - \nabla^2 \left(g^{\xi\eta} h_{\xi\eta}\right) \right)
\right\} \nn
&\nn
& - g_{\mu\nu}  \nabla^\rho \nabla^\sigma \left\{G'(R) R_{\rho\sigma} \left(
- h_{\xi\eta} R^{\xi\eta} + \nabla^\xi \nabla^\eta h_{\xi\eta}
  - \nabla^2 \left(g^{\xi\eta} h_{\xi\eta}\right) \right)
\right\}  \nn
&\nn
& - g_{\mu\nu} G(R) R_\rho^{\ \xi} R^{\rho\eta}h_{\xi\eta}
+ 2 G(R) R_\mu^{\ \xi} R_\nu^{\ \eta} h_{\xi\eta}
+ 2 G'(R) R_\rho^{\ \xi} R^{\rho\eta}h_{\xi\eta} R_{\mu\nu} \nn
&\nn
& + \left( \nabla_\mu \nabla_\nu - g_{\mu\nu} \nabla^2 \right) \left(
G'(R)  R_\rho^{\ \xi} R^{\rho\eta}h_{\xi\eta}  \right) \nn
&\nn
& + \frac{1}{2} \, g_{\mu\nu} G(R) R^{\rho\sigma}
\left\{\nabla^\xi\left(\nabla_\rho h_{\sigma\xi}
+ \nabla_\sigma h_{\rho\xi}\right) - \nabla^2 h_{\rho\sigma}
  - \nabla_\rho \nabla_\sigma \left(g^{\xi\eta} h_{\xi\eta}\right)\right\} \nn
&\nn
& - G(R) \left\{\nabla^\xi\left(\nabla_\rho h_{\mu\xi}
+ \nabla_\mu h_{\rho\xi}\right) - \nabla^2 h_{\rho\mu}
  - \nabla_\rho \nabla_\mu \left(g^{\xi\eta} h_{\xi\eta}\right)\right\}
R_\nu^{\ \rho} \nn
&\nn
& - G(R) \left\{\nabla^\xi\left(\nabla_\rho h_{\nu\xi}
+ \nabla_\nu h_{\rho\xi}\right) - \nabla^2 h_{\rho\nu}
  - \nabla_\rho \nabla_\nu \left(g^{\xi\eta} h_{\xi\eta}\right)\right\} R_\mu^{\ 
\rho} \nn
&\nn
& -  G'(R) R^{\rho\sigma} \left\{\nabla^\xi\left(\nabla_\rho
h_{\sigma\xi}
+ \nabla_\sigma h_{\rho\xi}\right) - \nabla^2 h_{\rho\sigma}
  - \nabla_\rho \nabla_\sigma \left(g^{\xi\eta} h_{\xi\eta}\right)\right\}
R_{\mu\nu}  \nn
&\nn
& + \left( \nabla_\mu \nabla_\nu - g_{\mu\nu} \nabla^2 \right) \left\{ G'(R) 
R^{\rho\sigma}  \left\{\nabla^\xi\left(\nabla_\rho h_{\sigma\xi}
+ \nabla_\sigma h_{\rho\xi}\right) - \nabla^2 h_{\rho\sigma}
  - \nabla_\rho \nabla_\sigma \left(g^{\xi\eta} h_{\xi\eta}\right)\right\} 
\right\} \nn
&\nn
& + \nabla_\mu \nabla^\rho \left\{ G(R) \left\{\nabla^\xi\left(\nabla_\rho 
h_{\nu\xi}
+ \nabla_\nu h_{\rho\xi}\right) - \nabla^2 h_{\rho\nu}
  - \nabla_\rho \nabla_\nu \left(g^{\xi\eta}
h_{\xi\eta}\right)\right\}\right\} \nn
&\nn
& + \nabla_\nu \nabla^\rho \left\{ G(R) \left\{\nabla^\xi\left(\nabla_\rho 
h_{\mu\xi}
+ \nabla_\mu h_{\rho\xi}\right) - \nabla^2 h_{\rho\mu}
  - \nabla_\rho \nabla_\nu \left(g^{\xi\eta} h_{\xi\eta}\right)\right\}
\right\} \nn
&\nn
& - \nabla^2 \left\{ G(R) \left\{\nabla^\xi\left(\nabla_\mu h_{\nu\xi}
+ \nabla_\nu h_{\mu\xi}\right) - \nabla^2 h_{\mu\nu}
  - \nabla_\mu \nabla_\nu \left(g^{\xi\eta} h_{\xi\eta}\right)\right\}
\right\} \nn
&\nn
& - g_{\mu\nu}  \nabla^\rho \nabla^\sigma \left\{ G(R) 
\left\{\nabla^\xi\left(\nabla_\rho h_{\sigma\xi}
+ \nabla_\sigma h_{\rho\xi}\right) - \nabla^2 h_{\rho\sigma}
  - \nabla_\rho \nabla_\sigma \left(g^{\xi\eta} h_{\xi\eta}\right)\right\}
\right\}  \nn
&\nn
& - \left( h_{\mu\nu} \nabla^2
  - g_{\mu\nu} h_{\xi\eta} \nabla^\xi \nabla^\eta \right) \left\{ F'(R) + G'(R) 
R_{\rho\sigma} R^{\rho\sigma}  \right\} \nonumber \\
& -  \frac{1}{2} \left( \delta_\mu^{\ \xi} \delta_\nu^{\ \eta}
  - g_{\mu\nu} g^{\xi\eta} \right) \left(
\nabla_\xi h_{\eta\lambda} + \nabla_\eta h_{\xi\lambda}
  - \nabla_\lambda h_{\xi\eta} \right) \partial^\lambda \left\{ F'(R) + G'(R) 
R_{\rho\sigma} R^{\rho\sigma}  \right\} \nn
&\nn
& - \frac{1}{2} \left\{
2 g^{\kappa\lambda}\left(
\nabla_\mu h_{\nu\lambda} + \nabla_\nu h_{\mu\lambda}
  - \nabla_\lambda h_{\mu\nu} \right) \nabla^\rho \left( G(R) 
R_{\rho\kappa}\right)
+ \nabla_\mu \left( h_{\xi\eta} \nabla^\xi \left( G(R) R^\eta_{\ \nu}\right) 
\right)
+ \nabla_\nu \left( h_{\xi\eta} \nabla^\xi \left( G(R) R^\eta_{\ \mu} \right) 
\right) \right. \nn
&\nn
& + \nabla_\mu \left( g^{\rho\sigma} \left( g^{\kappa\lambda}\left(
\nabla_\rho h_{\sigma\lambda} + \nabla_\sigma h_{\rho\lambda}
  - \nabla_\lambda h_{\sigma\rho} \right) \left( G(R) R_{\kappa\nu} \right)
+ g^{\kappa\lambda}\left(
\nabla_\rho h_{\nu\lambda} + \nabla_\nu h_{\rho\lambda}
  - \nabla_\lambda h_{\nu\rho} \right) \left( G(R) R_{\sigma\kappa} \right)
\right) \right)  \nn
&\nn
& + \nabla_\nu \left( g^{\rho\sigma} \left( g^{\kappa\lambda}\left(
\nabla_\rho h_{\sigma\lambda} + \nabla_\sigma h_{\rho\lambda}
  - \nabla_\lambda h g_{\sigma\rho} \right) \left( G(R) R_{\kappa\mu} \right)
+ g^{\kappa\lambda}\left(
\nabla_\rho h_{\mu\lambda} + \nabla_\mu h_{\rho\lambda}
  - \nabla_\lambda h_{\mu\rho} \right) \left( G(R) R_{\sigma\kappa}
\right)\right) \right) \nn
&\nn
& - 2 h_{\xi\eta} \nabla^\xi \nabla^\eta \left( G(R) R_{\mu\nu} \right)
  - g^{\xi\eta} \left( g^{\kappa\lambda}\left(
\nabla_\xi h_{\eta\lambda} + \nabla_\eta h_{\xi\lambda}
  - \nabla_\lambda h_{\xi\eta} \right) \nabla_\kappa \left( G(R) R_{\mu\nu}
\right) \right. \nn
&\nn
& \left. + g^{\kappa\lambda}\left(
\nabla_\xi h_{\mu\lambda} + \nabla_\mu h_{\xi\lambda}
  - \nabla_\lambda h_{\xi\mu} \right) \nabla_\eta \left( G(R) R_{\kappa\nu} 
\right)
+ g^{\kappa\lambda}\left(
\nabla_\xi h_{\nu\lambda} + \nabla_\nu h_{\xi\lambda}
  - \nabla_\lambda h_{\xi\nu} \right) \nabla_\eta \left( G(R) R_{\mu\kappa} 
\right)
\right) \nn
&\nn
&  - g^{\xi\eta} \nabla_\xi \left( g^{\kappa\lambda}\left(
\nabla_\eta h_{\mu\lambda} + \nabla_\mu h_{\eta\lambda}
  - \nabla_\lambda h_{\eta\mu} \right) \left( G(R) R_{\kappa\nu} \right)
+ g^{\kappa\lambda}\left(
\nabla_\eta h_{\nu\lambda} + \nabla_\nu h_{\eta\lambda}
  - \nabla_\lambda h_{\eta\nu} \right) \left( G(R) R_{\mu\kappa} \right)
\right) \nn
&\nn
& + 2 h_{\mu\nu}  \nabla^\rho \nabla^\sigma \left( G(R) R_{\rho\sigma} \right)
  - 2 g_{\mu\nu}  h_{\xi\eta} \nabla^\xi \nabla^\sigma \left( G(R) R^\eta_{\ 
\sigma} \right)
  - g_{\mu\nu}  g^{\tau\rho} g^{\kappa\lambda}\left(
\nabla_\tau h_{\rho\lambda} + \nabla_\rho h_{\tau\lambda}
  - \nabla_\lambda h_{\tau\rho} \right) \nabla^\sigma \left( G(R)
R_{\kappa\sigma} \right) \nn
&\nn
& \left. - g_{\mu\nu} \nabla^\rho \left( g^{\tau\sigma} \left( 
g^{\kappa\lambda}\left(
\nabla_\tau h_{\rho\lambda} + \nabla_\rho h_{\tau\lambda}
  - \nabla_\lambda h_{\tau\rho} \right) \left( G(R) R_{\kappa\sigma} \right)
+ g^{\kappa\lambda}\left(
\nabla_\tau h_{\sigma\lambda} + \nabla_\sigma h_{\tau\lambda}
  - \nabla_\lambda h_{\tau\sigma} \right) \left( G(R) R_{\rho\kappa}
\right) \right) \right) \right\} \nn
&\nn
& + \kappa^2 T_{\mu\nu}\, .
\end{align}
In the de Sitter background~(\ref{TuA5B}), Eq.~(\ref{TuA24B}) 
assumes the simplified form,
\begin{align}
0 = & \frac{1}{2} \, h_{\mu\nu} \left( F(R_0) + \frac{36}{l^4}
G(R_0)\right)
+ \frac{1}{2} \, g_{\mu\nu} F'(R_0)
\left( - \frac{3}{l^2} \, h + \nabla^\xi \nabla^\eta h_{\xi\eta}
  - \nabla^2 h \right) \nn
&\nn
& - \frac{1}{2} \left( F'(R_0) + \frac{36}{l^4} \, G'(R_0) + 
\frac{72}{l^2} \, G(R_0) \right)
\left\{\nabla^\rho\left(\nabla_\mu h_{\nu\rho}
+ \nabla_\nu h_{\mu\rho}\right) - \nabla^2 h_{\mu\nu}
  - \nabla_\mu \nabla_\nu \left(g^{\rho\lambda}
h_{\rho\lambda}\right)\right\}  \nn
&\nn
& + \left( F''(R_0) + \frac{36}{l^4} \, G''(R_0) \right)
\left( \nabla_\mu \nabla_\nu - g_{\mu\nu} \nabla^2
  - \frac{3}{l^2} \, g_{\mu\nu} \right) \left(
  - \frac{3}{l^2} \, h + \nabla^\xi \nabla^\eta h_{\xi\eta}
  - \nabla^2 h \right) \nn
&\nn
& + \frac{6}{l^2} \, G'(R_0) \left( \nabla_\mu \nabla_\nu - g_{\mu\nu} 
\nabla^2 \right)
\left(
  - \frac{3}{l^2} \, h + \nabla^\xi \nabla^\eta h_{\xi\eta}
  - \nabla^2 h \right)  \nn
&\nn
& - \frac{9}{l^4}\, g_{\mu\nu} G(R_0) h
+ \frac{18}{l^4} \, G(R_0) h_{\mu\nu}
+ \frac{54}{l^6} \, g_{\mu\nu} G'(R_0) h \nn
&\nn
& + \frac{9}{l^4} \, G'(R_0) \left( \nabla_\mu \nabla_\nu - g_{\mu\nu} 
\nabla^2 
\right) h
+ \frac{3}{l^2} \, G(R_0) g_{\mu\nu}  \left( 2 \nabla^\xi \nabla^\eta 
h_{\xi\eta}
  - \nabla^2 h  \right) \nn
&\nn
& - \frac{6}{l^2} \, G(R_0) \left\{\nabla^\xi\left(\nabla_\nu h_{\mu\xi}
+ \nabla_\mu h_{\nu\xi}\right) - \nabla^2 h_{\mu\nu}
  - \nabla_\mu \nabla_\nu h \right\} \nn
&\nn
& + \frac{6}{l^2} G'(R_0) \left( \nabla_\mu \nabla_\nu - g_{\mu\nu} \nabla^2
  - \frac{3}{l^2} \, g_{\mu\nu} \right)
\left( \nabla^\xi \nabla^\eta h_{\xi\eta} - \nabla^2 h \right) \nn
&\nn
& + G(R_0) \nabla_\mu \nabla^\rho \left\{ \left\{\nabla^\xi\left(\nabla_\rho 
h_{\nu\xi}
+ \nabla_\nu h_{\rho\xi}\right) - \nabla^2 h_{\rho\nu}
  - \nabla_\rho \nabla_\nu h \right\}\right\} \nn
&\nn
& + G(R_0) \nabla_\nu \nabla^\rho \left\{\nabla^\xi\left(\nabla_\rho h_{\mu\xi}
+ \nabla_\mu h_{\rho\xi}\right) - \nabla^2 h_{\rho\mu}
  - \nabla_\rho \nabla_\nu h \right\} \nn
&\nn
& - G(R_0) \nabla^2 \left\{\nabla^\xi\left(\nabla_\mu h_{\nu\xi}
+ \nabla_\nu h_{\mu\xi}\right) - \nabla^2 h_{\mu\nu}
  - \nabla_\mu \nabla_\nu h \right\} \nn
&\nn
& - G(R_0) g_{\mu\nu}  \nabla^\rho \nabla^\sigma
\left\{\nabla^\xi\left(\nabla_\rho h_{\sigma\xi}
+ \nabla_\sigma h_{\rho\xi}\right) - \nabla^2 h_{\rho\sigma}
  - \nabla_\rho \nabla_\sigma h \right\}  \nn
&\nn
& \left. + \frac{6}{l^2} \, G(R_0) \left( \nabla_\mu \nabla^\rho 
h_{\rho\nu}
+ \nabla_\nu \nabla^\rho h_{\rho\mu} - \nabla^2 h_{\mu\nu}
+ g_{\mu\nu} \nabla^\rho \nabla^\sigma h_{\rho\sigma} \right)
\right\} \nn
&\nn
& + \kappa^2 T_{\mu\nu} \, .  \label{TuA25B}
\end{align}
In the weak-field, slow-motion limit of GR the
linearized Einstein equations reduce to $\nabla^2 h_{\mu\nu}=-2\kappa^2
T_{\mu\nu}$ \cite{Wald:1984rg}. Hence, the coupling of the graviton to matter 
is given by the coefficient of $\nabla^2 h_{\mu\nu}$ 
in the linearized field equations~(\ref{TuA25B}) of our gravity model (where
we discard time derivatives and spatial derivatives of order higher than
second). The result is 
\begin{equation}
\label{TuA26}
\frac{1}{8\pi G_\mathrm{eff}} = \frac{1}{\kappa^2} \left[ F'(R_0) +
\frac{36}{l^2} \left( 2G(R_0) + \frac{G'(R_0)}{l^2} \right) \right] \, .
\end{equation}   
Then the turnaround radius~(\ref{TuA05}) is expressed as 
\begin{equation}
\label{TuA05BB}
r_\mathrm{TA}^3 = G_\mathrm{eff} M l^2
=\frac{\kappa^2 M l^2}{ 8\pi \left[ F'(R_0) + \frac{36}{l^2}\left(
\frac{G'(R_0) }{l^2}  + 2 G(R_0) \right) \right] } \, .
\end{equation}
As a partial consistency check of our equation~(\ref{TuA26}),
consider the
special  case of $F(R)=R^2$ gravity. While this model is a good
approximation of
Starobinsky inflation $F(R)=R+\mu R^2$ in the early, strongly
curved universe, it is well known that this theory (or, in $d$
spacetime dimensions, $F(R)=R^{d/2}$ \cite{Vollick:2007fh})  does not admit a
Newtonian limit \cite{PechlanerSexl} and suffers from other
problems as well \cite{Ferraris:1988zz,Sotiriou:2006hs,Nzioki:2009av}. 
Indeed, the exponent
$n$ of $F(R)=R^n $ gravity is severely constrained by the precession of
Mercury's perihelion to be \cite{Clifton:2005at,Clifton:2006kc,Barrow:2005dn,Zakharov:2006uq}
\be
n-1 = \left( 2.7 \pm 4.5 \right)\cdot 10^{-19} \,,
\ee
while the criterion $F''(R)\geq 0$ necessary to avoid the
notorious Dolgov-Kawasaki instability requires $n \geq 1$
\cite{Nojiri:2003ft,Faraoni:2006sy,Faraoni:2007yn}. For
the pathological model with $F(R)=R^2$ and $G(R)=0$, Eq.~(\ref{TuA26})
gives
\be
G_\mathrm{eff}= \frac{\kappa^2}{16\pi R_0} \,.
\ee
In order to take the Newtonian limit, one must be able to consider a
Minkowskian background, an assumption which complements our
Eq.~(\ref{TuA23}) and which is implemented when  $g_{\mu\nu}$ becomes the
Minkowski metric
$\eta_{\mu\nu}$. This Minkowski background can be seen as a de Sitter
space with zero curvature and $G_\mathrm{eff}$ diverges as $R_0 \rightarrow 0$,
which shows that the pathological theory $F(R)=R^2$ without Newtonian
limit leads to inconsistencies in our equations, as it should be. 

In the following sections we consider more concrete models.

\section{$R+R^2 +R_{\mu\nu} R^{\mu\nu}$ model}
\label{sec:4}
\setcounter{equation}{0}

Let us consider now the model \cite{Buchbinder:1992rb,Nojiri:2002qn},
\begin{equation}
\label{TuA1}
S = \frac{1}{2\kappa^2} \int d^4 x \, \sqrt{ - g } \Big[  R - \Lambda
+ a R^2 + b R_{\mu\nu} R^{\mu\nu} \Big] + S_\mathrm{matter} \, ,
\end{equation}
where $a$ and $b$ are constants and $S_\mathrm{matter}$ denotes the matter action.
This theory is known to be multiplicatively-renormalizable quantum gravity 
(for a review, see \cite{Buchbinder:1992rb}) which still has some unresolved issues with 
unitarity.

One could add to this Lagrangian density a
term proportional to the Kretschmann scalar $R_{\mu\nu\alpha\beta}
R^{\mu\nu\alpha\beta}$, but this does
not make the action more general. In fact, in four spacetime
dimensions, the integral of the Gauss-Bonnet  combination
\begin{equation}
\chi \equiv \int d^4 x \sqrt{-g} \left( R^2
-4R_{\mu\nu}R^{\mu\nu}+R_{ \mu\nu\alpha\beta}R^{ \mu\nu\alpha\beta}
\right) \, ,
\end{equation}
is a constant topological invariant, which allows one to eliminate the
Kretschmann term and reduce
the action
\begin{equation}
S'= \frac{1}{2\kappa^2} \int d^4 x \, \sqrt{ - g } \Big[  R - \Lambda
+ a R^2 + b R_{\mu\nu} R^{\mu\nu} +c
R_{\mu\nu\alpha\beta}R^{\mu\nu\alpha\beta} \Big]
+ S_\mathrm{matter} \,,
\end{equation}
where $c$ is another constant, to an action integral of the
form~(\ref{TuA1}) with new coefficients $a'=a-c $ and $ b'= b+4c$.

It is known  that the model~(\ref{TuA1}) contains a scalar mode and also a 
massive spin~2  ghost mode, in addition to the massless spin~2 mode, which is 
the usual graviton familiar from GR.
The existence of this ghost mode tells us that this model is not
unitary and is therefore inconsistent. This model is, however, regarded 
as a low-energy effective theory and, if we include the 
higher order corrections and non-perturbative effects, we may obtain 
a consistent theory.

Because
\begin{equation}
\label{TuA001}
F(R) = R - \Lambda + a R^2 \, , \quad \quad G(R) = b \, ,
\end{equation}
Eq.~(\ref{TuA22}) is reduced to 
\begin{equation}
\label{TuA6}
0 = \frac{3}{l^2} - \frac{\Lambda}{2}\, ,
\end{equation}
that is, if $ 1/l^2 \neq 0$,
\begin{equation}
\label{TuA7}
l^2 = \frac{6}{\Lambda} \, .
\end{equation}
Equation~(\ref{TuA26}) gives also  the effective gravitational coupling
$G_\mathrm{eff}$ as 
\begin{equation}
\label{TuA14}
8\pi G_\mathrm{eff} = \frac{\kappa^2}{ 1 +\frac{24}{l^2}\left( a+3b
\right) } \, ,
\end{equation}     
while the turnaround radius is 
\begin{equation}
\label{TuA15}
r_\mathrm{TA}^3 = \frac{3\kappa^2 \, M}{ 4\pi \Lambda  \left[ 1
+\frac{24}{l^2} \left( a +3b \right)  \right] } \, .
\end{equation}    
In Ref.~\cite{Capozziello:2018oiw}, it is required that the maximum
turnaround radius in any alternative theory of gravity 
be, at most, 10\% smaller than the corresponding radius~(\ref{TuA05})
in GR, 
\begin{equation}
\label{TuA16}
r_\mathrm{TA} \geq 0.9 \left( G_\mathrm{N}M l^2 \right)^{1/3}  \, .
\end{equation}
Applying this criterion here yields the constraint
\begin{equation}
\label{TuA17}
\frac{G_\mathrm{eff}}{R_\mathrm{dS}}
\geq \frac{0.182\, G_\mathrm{N}}{\Lambda} \, .
\end{equation}
Here $R_\mathrm{dS}$ is the scalar curvature of the geometry 
describing the de Sitter spacetime which solves 
the alternative theory of gravity and $\Lambda$ 
is the cosmological constant in GR, $\Lambda=3/l^2$ in the definition 
of~\cite{Capozziello:2018oiw}. 
In the case of the action~(\ref{TuA1}), Eq.~(\ref{TuA15}) 
in conjunction with the constraint~(\ref{TuA16}), yields 
\begin{equation}
\label{TuA18}
\frac{1}{1 +\frac{24}{l^2}\left(a+3b \right) } \geq 0.7\, ,
\end{equation}
which gives a constraint on the parameters $a$ and
$b$ in the model~(\ref{TuA1}).
We should note, however, that we have estimated the effective
gravitational coupling $G_\mathrm{eff}$ as 
given by the coupling of $h_{\mu\nu}$,  which may give Newton's law.
If we can know directly any parameter $\kappa^2$, $a$, or $b$ by any 
independent procedure, Eq.~(\ref{TuA18}) produces a more realistic
constraint on the model.  For example, it is not so clear whether 
the effective gravitational
coupling $G_\mathrm{eff}$ in the solution describing the Schwarzschild-de
Sitter spacetime (\ref{TuA8}) is identical with
$G_\mathrm{eff}$ in (\ref{TuA14}). The  effective gravitational coupling
$G_\mathrm{eff}$ in (\ref{TuA8}) depends on the definition of the mass $M$
in the modified gravity theory.

In the case of the critical gravity theory \cite{Lu:2011zk},
\begin{equation}
\label{TuA19}
a = - \frac{1}{\Lambda}= - \frac{l^2}{6}\, , \quad \quad
b = - 3 a = \frac{3}{\Lambda} = \frac{l^2}{2}\, ,
\end{equation}
the scalar mode does not propagate and the
massive spin two mode of the general $R+R^2+R_{\mu\nu} R^{\mu\nu}$
gravity becomes massless. Then, Eq.~(\ref{TuA19}) tells us that 
\begin{equation}
\label{TuA20}
\frac{1}{1 + \frac{24}{l^2}\left(a+3b\right)}=\frac{1}{33}\, ,
\quad
\quad 8\pi G_\mathrm{eff} = \frac{\kappa^2}{33} \, ,
\end{equation}
and therefore Eq.~(\ref{TuA18}) is not satisfied.

\section{A more general model}
\label{sec:5}
\setcounter{equation}{0}

Instead of the action (\ref{TuA1}), we consider the case that the
constants $\Lambda$, $a$, and $b$ depend on $R$,
\begin{equation}
\label{TuA021}
S = \int d^4 x \, \frac{\sqrt{ - g }}{2\kappa(R)^2} \Big[ R - \Lambda(R)
+ a(R) R^2 + b(R) R_{\mu\nu} R^{\mu\nu} \Big] + S_\mathrm{matter} \, ,
\end{equation}
that is,
\begin{equation}
\label{TuA022}
F(R) =  \frac{\kappa_0^2}{\kappa(R)^2} \left[  R - \Lambda(R) + a(R) R^2
\right] \, ,  \quad \quad G(R) = \frac{\kappa_0^2}{\kappa(R)^2} b(R) \, .
\end{equation}
(Similar to the previous section, adding a term
$cR_{\mu\nu\alpha\beta}R^{\mu\nu\alpha\beta}$ with constant coefficient
$c$ to the Lagrangian density does not add in generality.) 
We now write $\kappa$ in Eq.~(\ref{TuA21B}) as $\kappa_0$. In the
model consisting of one-loop corrected quantum $R^2$ gravity of 
Refs.~\cite{Buchbinder:1992rb,Myrzakulov:2014hca}, we have 
\begin{align}
\label{TuA023}
&  \kappa(R)^2 \sim \kappa_0^2 \left( 1 + \lambda_0 \beta_2 \tau
\right)^{0.77}\, , \quad \quad
  \kappa(R)^2 \Lambda(R) \sim \kappa_0^2 \Lambda_0 \left(
1 + \lambda_0 \beta_2 \tau \right)^{-0.55}\, , \\
&\nn
& \frac{a(R)}{ \kappa(R)^2} \sim \frac{a_0}{\kappa_0^2} \left( 1 +
\lambda_0 \beta_2 \tau \right) \, , \quad  \quad \frac{b(R)}{
\kappa(R)^2} \sim \frac{b_0}{ \kappa_0^2} \left( 1 +
\lambda_0 \beta_2 \tau \right) \, ,
\end{align}    
where
\begin{equation}
\label{TuA024}
\beta_2 = \frac{133}{10} \, , \quad \quad
\tau = \tau_1 \ln \left| \frac{R}{R_1} \right| \, .
\end{equation}
The $R$-dependent coefficients represent one-loop RG coupling constants.
The second interpretation of the same model is just a more complicated version 
of modified gravity which includes Ricci-squared term.

Therefore, we obtain
\begin{equation}
\label{TuA0025}
F(R) \sim R \left( 1 + \lambda_0 \beta_2 \tau \right)^{0.77}
  - \Lambda_0 \left( 1 + \lambda_0 \beta_2 \tau \right)^{-0.55}
+ a_0 R^2 \left( 1 + \lambda_0 \beta_2 \tau \right) \, , \quad
G(R) \sim b_0 \left( 1 + \lambda_0 \beta_2 \tau \right) \, .
\end{equation}
and Eq.~(\ref{TuA22}) then yields 
\begin{align}
\label{TuA025}
0 =& \frac{1}{2} \left[ R_0 \left( 1 + \lambda_0 \beta_2 \tau_0
\right)^{0.77}  - \Lambda_0 \left( 1 + \lambda_0 \beta_2 \tau_0
\right)^{-0.55} + a_0 R_0^2 \left( 1 + \lambda_0 \beta_2 \tau_0 \right)
\right] \nn
&\nn
& - \left. \frac{3}{l^2} \right[ \left( 1 + \lambda_0 \beta_2 \tau_0
\right)^{0.77} + 2 a_0 R_0 \left( 1 + \lambda_0 \beta_2 \tau_0 \right) \nn
&\nn
& \left. \qquad \qquad + 0.77 \lambda_0 \beta_2 \left( 1 + \lambda_0
\beta_2 \tau_0 \right)^{- 0.23} + 0.55 \, \frac{\Lambda_0 \lambda_0
\beta_2}{R_0} \left( 1 + \lambda_0 \beta_2 \tau_0 \right)^{-1.55}
+ a_0 \lambda_0 \beta_2 R_0 \right] \, ,
\end{align}
where
\begin{align}
\label{TuA026}
\tau_0 = \tau_1 \ln \left| \frac{R_0}{R_1} \right| \, .
\end{align}
It is difficult to solve Eq.~(\ref{TuA025}) but if we choose
$R_1=R_0=12/l^2$, that is, $\tau_0 = 0$, Eq.~(\ref{TuA025}) reduces to
\begin{equation}
\label{TuA027}
0 =\frac{3 - 0.77 \lambda_0 \beta_2}{l^2} - \Lambda_0\left(\frac{1}{2}
+ 0.14 \, \lambda_0 \beta_2 \right) - \frac{36a_0 \lambda_0
\beta_2}{l^4}\, ,
\end{equation}
which can be solved with respect to $l^2$, obtaining 
\begin{equation}
\label{TuA028}
l^2 = \frac{3 - 0.77 \lambda_0 \beta_2 \pm \sqrt{ \left( 3 - 0.77
\lambda_0 \beta_2 \right)^2 - 72 a_0 \lambda_0 \beta_2 \Lambda_0\left( 1
+ 0.28 \lambda_0 \beta_2 \right)}}
{2 \Lambda_0\left( 1 + 0.28 \lambda_0 \beta_2 \right)} \, .
\end{equation}
In Eq.~(\ref{TuA028}), the upper ($+$)  
sign corresponds to the classical limit~(\ref{TuA7}).  On the other hand,
Eq.~(\ref{TuA26}) gives 
\begin{align}
\label{TuA029}
\frac{1}{8\pi G_\mathrm{eff}} =& \frac{1}{\kappa^2} \left[
\left( 1 + \lambda_0 \beta_2 \tau_0 \right)^{0.77}
+ 2 a_0 R_0 \left( 1 + \lambda_0 \beta_2 \tau_0 \right) \right. \nn
&\nn
& \left. + 0.77 \lambda_0 \beta_2 \left( 1 + \lambda_0 \beta_2 \tau_0
\right)^{- 0.23} + 0.55 \, \frac{\Lambda_0 \lambda_0 \beta_2}{R_0} \left(
1 +
\lambda_0 \beta_2 \tau_0 \right)^{-1.55}  + a_0 \lambda_0 \beta_2 R_0 +
\frac{72}{l^2} b_0 \left( 1 + \lambda_0 \beta_2 \tau_0 \right) \right] \,.
\end{align}
Then, if we choose $R_1=R_0$, we find
\begin{equation}
\label{TuA030}
\frac{1}{8\pi G_\mathrm{eff}} = \frac{1}{\kappa^2} \left\{
1 + \frac{24 a_0}{l^2} + 0.77 \lambda_0 \beta_2
+ \frac{0.55 \Lambda_0 \lambda_0 \beta_2 l^2}{12}
+ \frac{12 a_0 \lambda_0 \beta_2}{l^2}
+ \frac{72}{l^2} b_0 \right\}
\end{equation}
and the turnaround radius is given by
\begin{equation}
\label{TuA031}
r_\mathrm{TA}^3 = \frac{8\pi \kappa^2 M l^2}{1 + \frac{24 a_0}{l^2} +
0.77 \lambda_0 \beta_2 + \frac{0.55 \Lambda_0 \lambda_0 \beta_2 l^2}{12}
+ \frac{12 a_0 \lambda_0 \beta_2}{l^2}
+ \frac{72}{l^2} b_0 } \, ,
\end{equation}
which provides the constraint, as in Eq.~(\ref{TuA18}),
\begin{equation}
\label{TuA032}
\frac{l^2}{1 + \frac{24 a_0}{l^2} + 0.77 \lambda_0 \beta_2
+ \frac{0.55 \Lambda_0 \lambda_0 \beta_2 l^2}{12}
+ \frac{12 a_0 \lambda_0 \beta_2}{l^2}
+ \frac{72}{l^2} b_0 }
\geq \frac{4.2}{\Lambda} \,.
\end{equation}
If we assume that the correction from Einstein gravity with a truly constant 
cosmological constant $\Lambda=\Lambda_0$, Eq.~(\ref{TuA028}) gives 
\begin{equation}
\label{TuA033}
l^2 = \frac{3}{\Lambda} \left( 1  - 0.31 \lambda_0 \beta_2 
 - 4 a_0 \lambda_0 \beta_2 \Lambda_0 \right) \, .
\end{equation}
Then combining with (\ref{TuA032}), we obtain the constraint 
\begin{equation}
\label{TuA34}
\frac{24 a_0}{l^2} + 1.08 \lambda_0 \beta_2
+ \frac{0.55 \Lambda_0 \lambda_0 \beta_2 l^2}{12}
+ 8 a_0 \lambda_0 \beta_2 \Lambda_0
+ 24\Lambda_0 b_0 \leq 0.4 \, ,
\end{equation}
as in (\ref{TuA18}).

\section{Discussion and conclusions}
\label{sec:6}
\setcounter{equation}{0}

Given the degeneracy between dark energy and modified gravity models
attempting to explain the present acceleration of the universe, and the
current level of theoretical and experimental effort aiming to detect and
study, or to constrain, possible deviations of gravity from Einstein's
theory \cite{Baker:2014zba,Berti:2015itd, PsaltisOzel,Psaltis:2014mca}, 
the turnaround
radius of large structures in cosmology could be very useful. Two
approaches to the turnaround radius in the context of GR
(\cite{Pavlidou:2013zha,Pavlidou:2014aia} and \cite{Faraoni:2015saa, Venice2017}) 
produce more or less the same numerical results. The second
approach, being based on the Hawking-Hayward quasilocal energy is
gauge-independent to any degree of approximation compatible with current
and foreseeable astronomical observations
\cite{Faraoni:2015saa,Venice2017}), but it becomes ill-defined in modified
gravity. For this reason, we used an alternative definition of turnaround
radius in our analysis in the context of modified gravity models.

Three previous works \cite{Faraoni:2015zqa, Capozziello:2018oiw,Lopes:2018uhq}
were restricted to scalar-tensor or $F(R)$ gravity (the latter is an
incarnation of the former class of theories). Here we discuss more general
classes of theories containing also the square of the Ricci tensor and
mixed terms. Allowing terms in $ R_{\mu\nu} R^{\mu\nu}$ to be present in
the action introduces extra degrees of freedom in comparison with pure
$F(R)$ or scalar-tensor gravity.

An important realization is that, even when the cosmic expansion is
identical in GR and in a modified gravity model, in general the time
dependence of the turnaround radius in the latter is different from
that of the corresponding turnaround radius in GR coupled with a perfect
fluid, because the effective gravitational coupling becomes
time-dependent.

To fix the ideas, we have imposed that the deviation of the turnaround
radius in modified gravity from its GR value is not larger than 10\% (this
figure may be debatable given the large error in the observational
determination of the turnaround radius \cite{Lee:2016oyu}, but it serves the
purpose of illustration). The constraint that we derive would already put
the critical gravity scenario of Ref.~\cite{Lu:2011zk} in jeopardy.
Similarly, more complicated models will be constrained by the turnaround
radius if and when reliable astronomical observations of this quantity
become available.

\begin{acknowledgments}

\noindent This work is supported in part by by MINECO (Spain), Project 
FIS2016-76363-P (S.D.O), by MEXT Grant-in-Aid for Scientific 
Research on Innovative Areas ``Cosmic Acceleration'' (No. 15H05890) 
and the JSPS Grant-in-Aid for Scientific
Research (C) No. 18K03615 (S.N.), and 
V.F. is supported by the Natural Sciences and Engineering Research
Council of  Canada (Grant No.~2016-03803).

\end{acknowledgments}


\begin{thebibliography}{99}

\bibitem{Buchbinder:1992rb}
I.~L.~Buchbinder, S.~D.~Odintsov, and I.~L.~Shapiro,
``Effective action in quantum gravity''  
(IOP: Bristol, UK, 1992). 

\bibitem{Callan:1985ia}
C.~G.~Callan, Jr., E.~J.~Martinec, M.~J.~Perry, and D.~Friedan,
Nucl.\ Phys.\ B {\bf 262} (1985) 593.
doi:10.1016/0550-3213(85)90506-1

\bibitem{Fradkin:1985ys}
E.~S.~Fradkin and A.~A.~Tseytlin,
Nucl.\ Phys.\ B {\bf 261} (1985) 1. 
Erratum: [Nucl.\ Phys.\ B {\bf 269} (1986) 745].
doi:10.1016/0550-3213(86)90522-5, 10.1016/0550-3213(85)90559-0

\bibitem{Capozziello:2011et}
S.~Capozziello and M.~De Laurentis,
Phys.\ Rept.\  {\bf 509} (2011) 167
doi:10.1016/j.physrep.2011.09.003
[arXiv:1108.6266 [gr-qc]].

\bibitem{Cai:2015emx}
Y.~F.~Cai, S.~Capozziello, M.~De Laurentis, and E.~N.~Saridakis,
Rept.\ Prog.\ Phys.\  {\bf 79} (2016) no.10,  106901
doi:10.1088/0034-4885/79/10/106901
[arXiv:1511.07586 [gr-qc]].

\bibitem{Nojiri:2006ri}
S.~Nojiri and S.~D.~Odintsov,
eConf C {\bf 0602061} (2006) 06
 [Int.\ J.\ Geom.\ Meth.\ Mod.\ Phys.\  {\bf 4} (2007) 115]
doi:10.1142/S0219887807001928 
[hep-th/0601213].

\bibitem{Sotiriou:2008rp}
T.~P.~Sotiriou and V.~Faraoni,
Rev.\ Mod.\ Phys.\  {\bf 82} (2010) 451 
doi:10.1103/RevModPhys.82.451 
[arXiv:0805.1726 [gr-qc]].

\bibitem{Nojiri:2010wj}
S.~Nojiri and S.~D.~Odintsov,
Phys.\ Rept.\  {\bf 505} (2011) 59 
doi:10.1016/j.physrep.2011.04.001 
[arXiv:1011.0544 [gr-qc]].

\bibitem{Nojiri:2017ncd}
S.~Nojiri, S.~D.~Odintsov, and V.~K.~Oikonomou,
Phys.\ Rept.\  {\bf 692} (2017) 1 
doi:10.1016/j.physrep.2017.06.001 
[arXiv:1705.11098 [gr-qc]].

\bibitem{Clifton:2011jh}
T.~Clifton, P.~G.~Ferreira, A.~Padilla, and C.~Skordis,
Phys.\ Rept.\  {\bf 513} (2012) 1 
doi:10.1016/j.physrep.2012.01.001 
[arXiv:1106.2476 [astro-ph.CO]].

\bibitem{Bamba:2012cp}
K.~Bamba, S.~Capozziello, S.~Nojiri, and S.~D.~Odintsov,
Astrophys.\ Space Sci.\  {\bf 342} (2012) 155
doi:10.1007/s10509-012-1181-8 
[arXiv:1205.3421 [gr-qc]].

\bibitem{Capozziello:2003tk}
S.~Capozziello, S.~Carloni, and A.~Troisi,
Recent Res.\ Dev.\ Astron.\ Astrophys.\  {\bf 1} (2003) 625
[astro-ph/0303041].

\bibitem{Nojiri:2003ft}
S.~Nojiri and S.~D.~Odintsov,
Phys.\ Rev.\ D {\bf 68} (2003) 123512 
doi:10.1103/PhysRevD.68.123512 
[hep-th/0307288].

\bibitem{Baker:2014zba}
T.~Baker, D.~Psaltis, and C.~Skordis,
Astrophys.\ J.\  {\bf 802} (2015) 63
doi:10.1088/0004-637X/802/1/63 
[arXiv:1412.3455 [astro-ph.CO]].

\bibitem{Stuchlik1} Z.~Stuchlik,  Bull. 
Astronomical Institutes of Czechoslovakia {\bf 34}, 129 (1983).

\bibitem{Stuchlik:1999qk}
Z.~Stuchlik and S.~Hledik, 
Phys.\ Rev.\ D {\bf 60} (1999) 044006. 
doi:10.1103/PhysRevD.60.044006

\bibitem{Stuchlik3} Z.~Stuchlik, P.~Slany, and S.~Hledik, 
Astron. Astrophys. {\bf 363},  425 (2000).

\bibitem{Stuchlik:2008dv}
Z.~Stuchlik,  
Mod.\ Phys.\ Lett.\ A {\bf 20} (2005) 561  
doi:10.1142/S0217732305016865 
[arXiv:0804.2266 [astro-ph]].

\bibitem{Mizony:2004sh}
M.~Mizony and M.~Lachieze-Rey,  
Astron.\ Astrophys.\  {\bf 434} (2005) 45 
doi:10.1051/0004-6361:20042195 
[gr-qc/0412084].

\bibitem{Stuchlik:2011zz}
Z.~Stuchlik and J.~Schee, 
JCAP {\bf 1109} (2011) 018. 
doi:10.1088/1475-7516/2011/09/018

\bibitem{Roupas:2013xga}
Z.~Roupas, M.~Axenides, G.~Georgiou, and E.~N.~Saridakis, 
Phys.\ Rev.\ D {\bf 89} (2014) no.8,  083002	
doi:10.1103/PhysRevD.89.083002	
[arXiv:1312.4893 [astro-ph.CO]].

\bibitem{Nolan:2014maa}
B.~C.~Nolan, 
Class.\ Quant.\ Grav.\  {\bf 31} (2014) no.23,  235008 
doi:10.1088/0264-9381/31/23/235008 
[arXiv:1408.0044 [gr-qc]].

\bibitem{Busha:2003sz}
M.~T.~Busha, F.~C.~Adams, R.~H.~Wechsler, and A.~E.~Evrard,
Astrophys.\ J.\  {\bf 596} (2003) 713 
doi:10.1086/378043 
[astro-ph/0305211].

\bibitem{Pavlidou:2013zha}
V.~Pavlidou and T.~N.~Tomaras, 
JCAP {\bf 1409} (2014) 020 
doi:10.1088/1475-7516/2014/09/020 
[arXiv:1310.1920 [astro-ph.CO]].

\bibitem{Pavlidou:2014aia}
V.~Pavlidou, N.~Tetradis, and T.~N.~Tomaras, 
JCAP {\bf 1405} (2014) 017 
doi:10.1088/1475-7516/2014/05/017 
[arXiv:1401.3742 [astro-ph.CO]].

\bibitem{Faraoni:2015zqa}
V.~Faraoni,  
Phys.\ Dark Univ.\  {\bf 11} (2016) 11 
doi:10.1016/j.dark.2015.11.001 
[arXiv:1508.00475 [gr-qc]].

\bibitem{Venice2017} 
V.~Faraoni, PoS EPS-HEP2017, 037,
European Physical Society conference on High Energy Physics 5-12 July
2017, Venice, Italy (2017)
[\href{https://pos.sissa.it/314/}\url{{https://pos.sissa.it/314/}}].

\bibitem{Hawking:1968qt}
S.~Hawking,
J.\ Math.\ Phys.\  {\bf 9} (1968) 598. 
doi:10.1063/1.1664615 

\bibitem{Hayward:1993ph}
S.~A.~Hayward,
Phys.\ Rev.\ D {\bf 49} (1994) 831
doi:10.1103/PhysRevD.49.831
[gr-qc/9303030].

\bibitem{Hayward:1994bu}
S.~A.~Hayward,
Phys.\ Rev.\ D {\bf 53} (1996) 1938
doi:10.1103/PhysRevD.53.1938
[gr-qc/9408002].

\bibitem{Faraoni:2015saa}
V.~Faraoni, M.~Lapierre-L\'{e}onard, and A.~Prain,
JCAP {\bf 1510},  013 (2015).
doi:10.1088/1475-7516/2015/10/013
[arXiv:1508.01725 [gr-qc]].

\bibitem{Szabados:2009eka}
L.~B.~Szabados,
Living Rev.\ Rel.\  {\bf 12} (2009) 4.
doi:10.12942/lrr-2009-4

\bibitem{Lapierre-Leonard:2017pvj}
M.~Lapierre-L\'{e}onard, V.~Faraoni, and F.~Hammad,
Phys.\ Rev.\ D {\bf 96} (2017) no.8,  083525
doi:10.1103/PhysRevD.96.083525
[arXiv:1710.06460 [gr-qc]].

\bibitem{Capozziello:2018oiw}
S.~Capozziello, K.~F.~Dialektopoulos, and O.~Luongo,
arXiv:1805.01233 [gr-qc].

\bibitem{Lopes:2018uhq}
R.~C.~C.~Lopes, R.~Voivodic, L.~R.~Abramo, and L.~Sodr\'e,
arXiv:1805.09918 [astro-ph.CO].


\bibitem{Bhattacharya:2016vur}
S.~Bhattacharya, K.~F.~Dialektopoulos, A.~E.~Romano, C.~Skordis and T.~N.~Tomaras,
JCAP {\bf 1707} (2017) no.07,  018
doi:10.1088/1475-7516/2017/07/018
[arXiv:1611.05055 [astro-ph.CO]].

\bibitem{Cai:2009qf}
R.~G.~Cai, L.~M.~Cao, Y.~P.~Hu, and N.~Ohta,
Phys.\ Rev.\ D {\bf 80} (2009) 104016 
doi:10.1103/PhysRevD.80.104016
[arXiv:0910.2387 [hep-th]].

\bibitem{Cai:2008mh}
R.~G.~Cai, L.~M.~Cao, Y.~P.~Hu, and S.~P.~Kim,
Phys.\ Rev.\ D {\bf 78} (2008) 124012
doi:10.1103/PhysRevD.78.124012
[arXiv:0810.2610 [hep-th]].

\bibitem{Wu:2007se}
S.~F.~Wu, B.~Wang, and G.~H.~Yang,
Nucl.\ Phys.\ B {\bf 799} (2008) 330
doi:10.1016/j.nuclphysb.2008.01.013
[arXiv:0711.1209 [hep-th]].

\bibitem{Cognola:2011nj}
G.~Cognola, O.~Gorbunova, L.~Sebastiani, and S.~Zerbini,
Phys.\ Rev.\ D {\bf 84} (2011) 023515
doi:10.1103/PhysRevD.84.023515
[arXiv:1104.2814 [gr-qc]].

\bibitem{Faraoni:2015sja}
V.~Faraoni,
Class.\ Quant.\ Grav.\  {\bf 33} (2016) no.1,  015007
doi:10.1088/0264-9381/33/1/015007 
[arXiv:1508.06849 [gr-qc]]. 

\bibitem{Hammad:2016yjq}
F.~Hammad,
Class.\ Quant.\ Grav.\  {\bf 33} (2016) no.23,  235016
doi:10.1088/0264-9381/33/23/235016
[arXiv:1611.03484 [gr-qc]].

\bibitem{Lee:2015upn}
J.~Lee, S.~Kim, and S.~C.~Rey,
Astrophys.\ J.\  {\bf 815} (2015) no.1,  43
doi:10.1088/0004-637X/815/1/43
[arXiv:1511.00056 [astro-ph.CO]].

\bibitem{Lee:2016qpt}
J.~Lee,
Astrophys.\ J.\  {\bf 832} (2016) no.2,  123
doi:10.3847/0004-637X/832/2/123
[arXiv:1603.06672 [astro-ph.CO]].

\bibitem{Lee:2016oyu}
J.~Lee and G.~Yepes,
Astrophys.\ J.\  {\bf 832} (2016) no.2,  185
doi:10.3847/0004-637X/832/2/185
[arXiv:1608.01422 [astro-ph.CO]].



\bibitem{Lee:2016bec} 
J.~Lee and B.~Li,
Astrophys.\ J.\  {\bf 842}, no. 1, 2 (2017)
doi:10.3847/1538-4357/aa706f
[arXiv:1610.07268 [astro-ph.CO]].


\bibitem{Lee:2017ejv} 
J.~Lee,
Astrophys.\ J.\  {\bf 856}, no. 1, 57 (2018)
doi:10.3847/1538-4357/aab358
[arXiv:1709.06903 [astro-ph.CO]].



\bibitem{Wald:1984rg}
R.~M.~Wald, ``General Relativity'' (Chicago University Press, 
Chicago, 1984)  
doi:10.7208/chicago/9780226870373.001.0001

\bibitem{Caldwell:2003vq}
R.~R.~Caldwell, M.~Kamionkowski, and N.~N.~Weinberg,
Phys.\ Rev.\ Lett.\  {\bf 91} (2003) 071301,
doi:10.1103/PhysRevLett.91.071301 [astro-ph/0302506].

\bibitem{Frampton:2011sp}
P.~H.~Frampton, K.~J.~Ludwick, and R.~J.~Scherrer,
Phys.\ Rev.\ D {\bf 84} (2011) 063003
doi:10.1103/PhysRevD.84.063003
[arXiv:1106.4996 [astro-ph.CO]].

\bibitem{Frampton:2011rh}
P.~H.~Frampton, K.~J.~Ludwick, S.~Nojiri, S.~D.~Odintsov, and
R.~J.~Scherrer,
Phys.\ Lett.\ B {\bf 708} (2012) 204
doi:10.1016/j.physletb.2012.01.048
[arXiv:1108.0067 [hep-th]].

\bibitem{Vollick:2007fh}
D.~N.~Vollick,
Phys.\ Rev.\ D {\bf 76} (2007) 124001
doi:10.1103/PhysRevD.76.124001
[arXiv:0710.1859 [gr-qc]].

\bibitem{PechlanerSexl} 
E.~Pechlaner and R.~Sexl, Commun. Math. Phys. {\bf
2}, 165 (1966).

\bibitem{Ferraris:1988zz}
M.~Ferraris, M.~Francaviglia, and G.~Magnano,
Class.\ Quant.\ Grav.\  {\bf 5} (1988) L95.
doi:10.1088/0264-9381/5/6/002

\bibitem{Sotiriou:2006hs}
T.~P.~Sotiriou,
Class.\ Quant.\ Grav.\  {\bf 23} (2006) 5117
doi:10.1088/0264-9381/23/17/003
[gr-qc/0604028].

\bibitem{Nzioki:2009av}
A.~M.~Nzioki, S.~Carloni, R.~Goswami, and P.~K.~S.~Dunsby,
Phys.\ Rev.\ D {\bf 81} (2010) 084028
doi:10.1103/PhysRevD.81.084028
[arXiv:0908.3333 [gr-qc]].

\bibitem{Clifton:2005at}
T.~Clifton and J.~D.~Barrow,
Phys.\ Rev.\ D {\bf 72} (2005) 123003
doi:10.1103/PhysRevD.72.123003
[gr-qc/0511076].

\bibitem{Clifton:2006kc}
T.~Clifton and J.~D.~Barrow,
Class.\ Quant.\ Grav.\  {\bf 23} (2006) 2951
doi:10.1088/0264-9381/23/9/011
[gr-qc/0601118].

\bibitem{Barrow:2005dn}
J.~D.~Barrow and T.~Clifton,
Class.\ Quant.\ Grav.\  {\bf 23} (2006) L1
doi:10.1088/0264-9381/23/1/L01
[gr-qc/0509085].

\bibitem{Zakharov:2006uq}
A.~F.~Zakharov, A.~A.~Nucita, F.~De Paolis, and G.~Ingrosso,
Phys.\ Rev.\ D {\bf 74} (2006) 107101
doi:10.1103/PhysRevD.74.107101
[astro-ph/0611051].

\bibitem{Faraoni:2006sy}
V.~Faraoni,
Phys.\ Rev.\ D {\bf 74} (2006) 104017
doi:10.1103/PhysRevD.74.104017
[astro-ph/0610734].

\bibitem{Faraoni:2007yn}
V.~Faraoni,
Phys.\ Rev.\ D {\bf 75} (2007) 067302
doi:10.1103/PhysRevD.75.067302
[gr-qc/0703044 [GR-QC]].

\bibitem{Nojiri:2002qn}
S.~Nojiri and S.~D.~Odintsov,
Phys.\ Rev.\ D {\bf 66} (2002) 044012
doi:10.1103/PhysRevD.66.044012
[hep-th/0204112].

\bibitem{Lu:2011zk}
H.~Lu and C.~N.~Pope,
Phys.\ Rev.\ Lett.\  {\bf 106} (2011) 181302
doi:10.1103/PhysRevLett.106.181302
[arXiv:1101.1971 [hep-th]].

\bibitem{Myrzakulov:2014hca}
R.~Myrzakulov, S.~Odintsov, and L.~Sebastiani,
Phys.\ Rev.\ D {\bf 91} (2015) no.8,  083529
doi:10.1103/PhysRevD.91.083529
[arXiv:1412.1073 [gr-qc]].

\bibitem{Berti:2015itd}
E.~Berti {\it et al.},
Class.\ Quant.\ Grav.\  {\bf 32} (2015) 243001
doi:10.1088/0264-9381/32/24/243001
[arXiv:1501.07274 [gr-qc]].

\bibitem{PsaltisOzel} 
D.~Psaltis and F.~\"Ozel, Physics Today {\bf 71} (4)
70 (2018). 

\bibitem{Psaltis:2014mca}
D.~Psaltis, F.~\"{O}zel, C.~K.~Chan, and D.~P.~Marrone,
Astrophys.\ J.\  {\bf 814} (2015) no.~2,  115
doi:10.1088/0004-637X/814/2/115
[arXiv:1411.1454 [astro-ph.HE]].



\end{thebibliography}
\end{document}